\documentclass[letterpaper, 10 pt, conference]{ieeeconf}  

\IEEEoverridecommandlockouts                              

\usepackage{graphicx}
\usepackage{xcolor}
\usepackage{comment}
\usepackage{amsmath,amssymb,amsfonts}

\usepackage{amsthm}
\usepackage{algorithmic}
\usepackage{graphicx}
\usepackage{textcomp}
\usepackage{comment}
\usepackage{mathtools, cuted}
\usepackage{relsize}
\usepackage{comment}
\usepackage{cite}

\usepackage{algorithm}
\usepackage{multicol}
\usepackage[normalem]{ulem}
\usepackage{pgfplots}
\usepackage{booktabs}

\pgfplotsset{compat=newest}
\newlength\figureheight
\newlength\figurewidth
\newtheorem{remark}{Remark}
\newtheorem{lemma}{Lemma}
\newtheorem{theorem}{Theorem}
\newtheorem{definition}{Definition}

\title{\LARGE \bf
Identification of contractive Lur’e-type systems via kernel-based Lipschitz design
}

\def\CD#1{\textcolor{red}{#1}}
\def\CDnote#1{\textcolor{red}{\textbf{[CD $\bullet$} \textit{#1}\textbf{]}}}

\author{Cesare Donati${}^{1,2}$,
Fabrizio Dabbene${}^{2}$, Constantino Lagoa${}^{3}$, Carlo Novara${}^{1}$, and Yoshio Ebihara${}^{4}$
\thanks{This work was supported by Japan Science
and Technology Agency (JST) as part of Adopting Sustainable
Partnerships for Innovative Research Ecosystem (ASPIRE),
Grant Number JPMJAP2402.}
\thanks{${}^{1}$DET, Politecnico di Torino, Corso Duca degli Abruzzi 24, Torino, Italy. {\tt\small cesare.donati, carlo.novara@polito.it}}
\thanks{${}^{2}$CNR-IEIIT, c/o Politecnico di Torino, Corso Duca degli Abruzzi 24, Torino, Italy.
{\tt\small fabrizio.dabbene@cnr.it}}
\thanks{${}^{3}$EECS, The Pennsylvania State University, University Park, PA, USA.
{\tt\small cml18@psu.edu}}
\thanks{${}^{4}$Faculty of Information Science and Electrical Engineering,
Kyushu University, 744 Motooka, Nishi-ku, Fukuoka 819-0395, Japan. {\tt\small ebihara@ees.kyushu-u.ac.jp}}%
}

\begin{document}

\maketitle
\thispagestyle{empty}
\pagestyle{empty}

\begin{abstract}
This paper addresses the problem of identifying contractive Lur’e-type systems. Specifically, it proposes an identification framework that integrates linear prior knowledge with a kernel representation of the nonlinear feedback while systematically enforcing contractivity via Lipschitz constant design. The resulting algorithms provide models that are accurate in prediction, interpretable, and faithful to the contractive nature of the true system. Numerical experiments demonstrate that enforcing contractivity significantly improves parameter estimation and yields models that are both accurate and physically meaningful.
\end{abstract}


\section{Introduction}
The identification of nonlinear dynamical systems is a longstanding challenge in control and systems theory \cite{schoukens2019nonlinear,ribeiro2020smoothness}, with applications ranging from vehicle dynamics and aerospace engineering to robotics and epidemiology \cite{ljung2010perspectives}. In many scenarios, purely black-box models such as neural networks can provide accurate predictions but often lack interpretability and guarantees on stability \cite{dong2023neural}. Conversely, purely physics-based models are interpretable but may fail to capture unmodeled effects or nonlinearities present in real-world data \cite{donati2025combining}. Bridging this gap requires frameworks that combine prior physical knowledge with data-driven components, while at the same time enforcing structural properties that guarantee meaningful and reliable dynamical behavior \cite{donati2025combining}.  

A particularly relevant class of nonlinear systems is represented by \emph{Lur’e-type systems}, consisting of a linear dynamical part in feedback with a static nonlinearity. This class of systems is of significant practical interest, as many real-world engineering processes can be effectively represented within this framework (see, e.g., \cite{zames1968stability,zemouche2013lmi,REINDERS2021104660,
giaccagli2023lmi}).
Within this class, \emph{contractive Lur’e systems} \cite{shima_contractivity_2025} are of special interest. Contractivity ensures exponential convergence between arbitrary trajectories of nonlinear systems \cite{FB-CTDS}. 
Moreover, it offers substantial benefits for analyzing nonlinear dynamical phenomena, including synchronization in networked systems, the design of distributed controllers, and improved robustness against noise and disturbances (see \cite{shima_contractivity_2025} and references therein). 

When contractivity is an intrinsic property of the true system, it is not automatically preserved in the identification process. In fact, standard identification methods may yield models that fit the observed trajectories well but fail to inherit the contractive nature of the underlying dynamics.
Embedding this property into system identification procedures can provide models that are not only accurate in terms of prediction and parameter estimation but also faithful to the underlying system's fundamental dynamical structure.  
Within this context, while several works have addressed the identification of stable Lur’e-type systems (e.g., \cite{shakib2022computationally,richards2024output}), to the best of our knowledge, no existing approach simultaneously enforces contractivity and compensates for unmodeled dynamics within the system model, exploiting system physical knowledge.

To this aim, recent advances in \emph{kernel-based methods} \cite{aronszajn1950theory,scholkopf2001.reprthm,wahba1990spline} have opened new perspectives for nonlinear system identification \cite{donati2025kernelTAC}. Kernels offer a flexible yet theoretically grounded framework for function approximation, allowing residual dynamics to be modeled in Reproducing Kernel Hilbert Spaces (RKHSs). This allows parametric models to be complemented with data-driven corrections in a parsimonious way, while retaining interpretability. 
Moreover, such approaches are becoming increasingly adopted in learning problems, as they enable the systematic enforcement of desired properties in the learned models through the design of the kernel function or the choice of the regularization term \cite{scandella2024kernel}.
For instance, recent theoretical results have shown that the Lipschitz properties of kernel-based models can be explicitly characterized and controlled \cite{van2022kernel}, making them suitable for enforcing contractivity conditions. 

In this paper, we propose a {kernel-based identification framework tailored to contractive Lur’e-type systems}. Building on earlier work on physics-informed system identification \cite{donati2025combining,donati2025kernelTAC}, our method integrates prior knowledge of the linear structure with flexible kernel-based models of the nonlinear feedback. Crucially, we introduce a mechanism to enforce contractivity by jointly tuning the model parameters and the kernel regularization, thereby ensuring that the identified model preserves the contractivity properties of the true system. We show that this approach not only provides accurate predictions but also improves the estimation of physical parameters by mitigating the bias induced by noise and unmodeled effects.  

The main contributions of this work are as follows: (i) We develop a kernel-based identification framework for Lur’e-type systems that explicitly enforces contractivity; (ii) We derive conditions under which the kernel-based residual dynamics remain Lipschitz continuous with a given contraction factor; (iii) We propose two practical yet effective algorithms for identifying contractive models, balancing prediction accuracy, model complexity, and physical consistency; (iv) Through numerical experiments, we demonstrate that enforcing contractivity yields more reliable parameter estimates, improving the physical accuracy of the identified models.  

\subsubsection*{Outline}  
The remainder of the paper is organized as follows. Section~\ref{sec:problem-formulation} describes the general identification framework combining structured physical knowledge with residual models. In Section~\ref{sec:spec.case}, we specialize this framework to contractive Lur’e-type systems, while in Section~\ref{sec:lipdesign} we derive the conditions required to guarantee contractivity and discuss the Lipschitz-based design of kernel residuals. Section~\ref{sec:prop_approach} presents the proposed identification algorithms and the procedure for enforcing contractivity. Section~\ref{sec:examples} provides numerical experiments illustrating the effectiveness of the method. Finally, Section~\ref{sec:conclusions} concludes the paper and outlines directions for future research.

\subsubsection*{Notation}
We denote by $\mathbb{S}$$_n^{+}$ the set of positive definite matrices of size $n \times n$. Given a vector $v\in\mathbb R$$^{n}$ and $P \in \mathbb{S}$$_n^{+}$, we denote the weighted Euclidean norm $\sqrt{v^\top P v}$ by $\|v\|_{P}$. For a matrix 
$M$, $\|M\|_2$ denotes the induced matrix $\ell_2$-norm. $I_N$ is the identity matrix in $\mathbb{R}$$^{N,N}$.


\section{Identification framework}\label{sec:problem-formulation}
Let us consider a nonlinear, discrete-time dynamical system of the form
\begin{equation}
\begin{aligned}
    x_{t+1} &= f(x_t,u_t,\theta) + \Delta(x_t,u_t) + v_t,\\
    y_t &= h(x_t,u_t,\theta)+w_t,
\end{aligned}
    \label{eqn:nl.sys}
\end{equation}
where $x_t \in \mathcal{X} \subset 
{\mathbb{R}}$$^{n_x}$, $u_t\in\mathcal{U} \subset  
\mathbb R$$^{n_u}$, $y_t\in\mathcal{Y} \subset  
\mathbb R$$^{n_y}$ denotes the system state, a measured input, and the observed output at time $t$. Here, $\mathcal{X}$, $\mathcal{U}$, and $\mathcal{Y}$ denote the state, input, and output domains, respectively, within which the system is to be identified.
Disturbance terms $v_t$ and $w_t$ denote process and measurement noise, respectively.
The functions $f,h$ represent a known model structure (e.g., from physics) that captures the nominal dynamics of the system, parameterized by an unknown but fixed vector $\theta\in\Theta\subset\mathbb{R}$$^{n_\theta}$. These terms are intended to capture the relevant dynamics of the system, without trying to explain details that may not be well-known. Conversely, the term $\Delta$ accounts for unmodeled dynamics or residual effects not captured by the parametric model $f$.
A dataset of measured input-output pairs, $\mathcal{D}$ $= \{(u_1, y_1), \dots, (u_T, y_T)\}$, collected from observations of system \eqref{eqn:nl.sys}, is available.

The goal is to identify a predictive model to be used for approximating the system dynamics \eqref{eqn:nl.sys}. Within this framework, the predictive model is defined by the following structure
\begin{equation}
\begin{aligned}
    \hat{x}_{t+1} &= f(\hat x_t, u_t, \hat{\theta}) + \delta(\hat x_t, u_t, \omega),\\
    \hat{y}_{t} &= h(\hat x_t, u_t, \hat{\theta}),
\end{aligned}
    \label{eqn:model}
\end{equation}
where $\hat x_t, \hat y_t$ denote the predicted state and the predicted output, $\hat{\theta}$ is an estimate of the true parameter vector $\theta$, and $ \delta $ is a learned or approximated function intended to capture the unmodeled dynamics represented by $ \Delta $. The functions $ f, h$ share the same known parametric form as in the true system but are evaluated at the estimated parameters and state. Here, we note that the function $ \delta $ may be implemented using a flexible model class, such as weighted basis function dictionaries {\cite{donati2025combining}}, kernel-based functions {\cite{donati2025kernelTAC}}, or neural networks {\cite{dong2023neural}}. The vector $ \omega \in\Omega\subset\mathbb{R}$$^{n_\omega} $ denotes the parameters of the function approximator used to model the residual dynamics. 
This modeling framework decomposes the system dynamics into two components: structured terms that incorporate prior knowledge through $ f, h$, and a data-driven correction term $ \delta $, enabling both interpretability and adaptability. Specifically, the proposed framework focuses on models with the structure \eqref{eqn:model}, 
where
the residual $\delta$ is identified according to some parsimony criterion (that will be specified later), to minimize its complexity \cite{donati2025combining}.
This class of models is closely related to the identification framework proposed in \cite{donati2025combining} and~{\cite{donati2025kernelTAC}, where similar decompositions were shown to enable accurate system identification. In particular, those works demonstrate that effective compensation of the residual term $ \Delta $ through the learned correction $ \delta $ can lead to improved estimation accuracy of the parametric component $ \hat{\theta} $.

\section{Contractive Lur'e-type systems}\label{sec:spec.case}

In this paper, we specialize on a relevant subclass of nonlinear systems known as {contractive Lur'e-type systems}, which naturally arise in many engineering applications~\cite{giaccagli2023lmi}. 

Lur'e-type systems are characterized by a linear dynamical part in feedback with a static nonlinearity, and can be obtained as a specialization of the general model in~\eqref{eqn:nl.sys}, yielding the following nominal form
\begin{equation}
\begin{aligned}
x_{t+1} &= A(\theta) x_t + B(\theta)u_t+{F}\phi(Cx_t) + v_t,\\
y_t &= C x_t + w_t.
\end{aligned}
\label{eqn:lure.sys}
\end{equation}
Here, $\phi: \mathcal{X} \to \mathbb R$$^{n_\phi}$ is an unknown static nonlinear function, while the matrices $A\in\mathbb R$$^{n_x,n_x}$ and $B\in\mathbb R$$^{n_x,n_u}$,
parametrized in $\theta$, define the linear dynamics. The output is given by a linear mapping of the state through the known matrix $C \in \mathbb{R}$$^{n_y, n_x}$. On the other hand, the known matrix $F \in \mathbb{R}$$^{n_x, n_\phi}$ scales the contribution of the nonlinearity and allows selecting which state components are affected by it, if such structural information is available. In particular, if all the states are affected, one can take $n_\phi = n_x$ and $F = I_{n_x}$.

Within this work, we restrict our attention to the identification of systems that satisfy the property of contractivity, formally defined next.
\begin{definition}[Contractivity \cite{shima_contractivity_2025}]\label{def:contract}\itshape
Consider $P \in \mathbb{S}$$_n^{+}$ and a constant $0 \leq \ell < 1$. 
System \eqref{eqn:lure.sys} is said to be \emph{strongly contracting} with respect to the norm $\|\cdot\|_{P}$ and contraction factor $\ell$ if, for any pair of solutions 
$ x_{t+\tau}$ and $ x'_{t+\tau}$ of \eqref{eqn:lure.sys} under zero input and no process noise, i.e., $u_t\equiv0$, $v_t\equiv0$, $\forall t$, the following inequality holds for all $\tau\geq 0$:
$$
  \| x_{t+\tau} -  x'_{t+\tau}\|_{P} \leq \ell^{\tau} \| x_t -  x'_t\|_{P}.
$$
\end{definition}

Note that, considering the case $\tau = 1$ in Definition~\ref{def:contract}, we have that system~\eqref{eqn:lure.sys} needs to be Lipschitz continuous with Lipschitz constant $0 \leq \ell < 1$ to be contractive.

\begin{remark}[Contractivity in the Euclidean norm]\itshape~We note that the choice of $P$ can itself affect the identification procedure. In the remainder of the paper, we will focus on the case where $P = I_{n_x}$, i.e., contractivity is considered with respect to the standard Euclidean norm $\|\cdot\|_2$.  This aspect will be extended in future work, where the selection of $P$ will also influence the contraction properties of the identified model.
\end{remark}

Thus, in the following, we propose an effective framework for the identification of contractive Lur'e-type systems. 
Let us consider \eqref{eqn:nl.sys} and \eqref{eqn:lure.sys}. The known structure, $ f $, corresponds to the linear part of the Lur’e model, i.e., $f(x_t,u_t,\theta) \doteq A(\theta)x_t+ B(\theta)u_t$.
Similarly, we have $h(x_t,u_t,\theta) \doteq Cx_t$. Thus, the residual term, $ \delta(x_t, \omega), $ accounts for the unknown feedback nonlinearity $\phi(Cx_t) $. 
%
%
%

To approximate this nonlinear term, we adopt a kernel-based identification strategy based on~\cite{donati2025kernelTAC}, which provides a structured framework to integrate prior physical knowledge with flexible kernel-based models. 
Building on this framework, and inspired by recent insights from~\cite{van2022kernel}, we propose an identification method that ensures that the resulting model satisfies a Lipschitz condition with a specifically designed constant. In this way, the identified model not only yields accurate estimates of the parameters characterizing \eqref{eqn:lure.sys} and low prediction errors, but also preserves the intrinsic contractivity property of the underlying system.

In what follows, we provide an overview of kernels and the kernel-based modeling framework introduced in~\cite{donati2025kernelTAC}, specializing it to the class of Lur’e-type systems. We then present our approach for enforcing contractivity in the identified model.

\subsection{Kernel-based modeling framework for Lur’e systems}
\subsubsection{Kernel-based approximation}\label{ssec:KernelTheory}
Let us first recall the core elements of kernel-based nonlinear function approximation \cite{scholkopf2001.reprthm,aronszajn1950theory,wahba1990spline}, which support the modeling framework adopted in \cite{donati2025kernelTAC} and in this section. A \emph{positive definite kernel} is a symmetric function $\kappa: \mathcal{Z} , \mathcal{Z} \to \mathbb{R}$ such that for any finite set $\{z_1, \dots, z_n\} \subset \mathcal{Z}$, $\sum_{i=1}^n\sum_{j=1}^n c_ic_j\kappa(z_i,z_j)\ge0$ holds for all $c_1,\dots,c_n\in\mathbb R$.
Any such kernel induces a unique RKHS $\mathcal{H}$, a Hilbert space of functions where for any $z\in\mathcal{Z}$, $\kappa(\cdot, z)\in\mathcal{H}$ and the \emph{reproducing property} holds: for all $h \in \mathcal{H}$ and $z \in \mathcal{Z}$, $\langle h(\cdot), \kappa(\cdot, z) \rangle_{\mathcal{H}} = h(z)$.

Given a dataset $\mathcal{D} $=$ \{(z_1, \xi_1),\dots,(z_T, \xi_T)\}$, where the unknown function $g \in \mathcal{H}$ relates $z$ and $\xi$ via 
\begin{equation}
\xi = g(z) + d,
\label{eqn:sample.sys}
\end{equation} and $d$ represents noise, an estimate $\hat g$ is obtained by solving the regularized problem
\begin{equation}
\hat g = \arg \min_{g \in \mathcal{H}} \sum_{t=1}^T [\xi_t - g(z_t)]^2 + \gamma \|g\|_{\mathcal{H}}^2,
\label{eqn:reg.ls.prob.kernel}
\end{equation}
where $\gamma > 0$ is a regularization parameter {and $\|g\|_{\mathcal{H}}=\sqrt{\langle g,g\rangle}_\mathcal{H}$ is the norm in $\mathcal{H}$, introduced by the inner product $\langle \cdot,\cdot\rangle_\mathcal{H}$. The \emph{representer theorem} \cite{scholkopf2001.reprthm} ensures that the solution admits a closed-form finite expansion:
$$
\hat g(z) = \sum_{j=1}^T \omega_j \kappa(z, z_j), \quad \omega = (K + \gamma I_T)^{-1} Y,
$$
where $K$ is the positive semidefinite kernel matrix with entries $K_{ij} = \kappa(z_i, z_j)$ and $Y = [\xi_1, \dots, \xi_T]^\top$. This result enables flexible function learning while controlling smoothness and complexity through the norm in $\mathcal{H}$ and~$\gamma$. The reader is referred to~\cite{scholkopf2001.reprthm,aronszajn1950theory,wahba1990spline} for further details on kernel-based approximation theory and its application to system identification, and to \cite{micchelli2005learning,alvarez2012kernels} for extensions to vector-valued kernels.

\subsubsection{Kernel methods with structural priors \cite{donati2025kernelTAC}}\label{sec:lure.kernel.method}  
Let us now consider the framework in \cite{donati2025kernelTAC}, and the Lur'e-type nonlinear system of the form \eqref{eqn:lure.sys}. Here, $A(\theta)$, $B(\theta)$, and $C$ represent the known linear structure and are functions of the unknown parameter vector $\theta$, while the nonlinear feedback term $\phi(Cx_t)$ is unknown and assumed to lie in an RKHS $\mathcal{H}$ associated with
a chosen kernel $\kappa$. Without loss of generality, we restrict our attention to the scalar case $n_x=1$, $n_u=1$, so that both $A(\theta)$ and $B(\theta)$ reduce to scalar functions of $\theta$. Moreover, we initially further assume $C=I_{n_x}$. The extension to the general multi-dimensional setting is straightforward and will be discussed in the following part of this section.

Given a dataset $\mathcal{D}$ $= \{ (u_0,x_0),\dots , (u_{T-1},x_{T-1})\} \cup \{x_T\}$, the identification task can be cast as the minimization of the following regularized cost function:
\begin{equation}
    J(\theta,\delta) \doteq \sum_{t=0}^{T-1} \left[ x_{t+1} {-} A(\theta)x_{t} {-} B(\theta)u_{t} {-} \delta(x_t) \right]^2 
    + \gamma \|\delta\|_{\mathcal{H}}^2,
    \label{eq:kernel.cost}
\end{equation}
where $\gamma > 0$ balances model fit and complexity. In particular, the regularization term $\|\delta\|_{\mathcal{H}}^2$ enforces parsimony of the learned residual dynamics $\delta$. 
The identification problem thus consists in jointly estimating the parametric component $\theta$ and the residual model $\delta$ by solving:
\begin{equation}
(\theta^\star, \delta^\star) = \arg\min_{\substack{\theta \in \Theta\\\delta \in \mathcal{H}}} J(\theta,\delta).
\label{eq:kernel.problem}
\end{equation}

In this setting, the kernel-based approximation framework of Section~\ref{ssec:KernelTheory} applies considering input $(x_t,u_t)$, and output $x_{t+1}$, for $t = 0, \dots, T-1$.

Let $K \in {\mathbb{R}}^{T , T}$ denote the kernel matrix with entries $K_{ij} = \kappa(x_i, x_j)$, and define the vectors
\begin{equation}
\begin{gathered}
    X {=} [x_1, \dots, x_T]^\top, \\
    \Gamma(\theta) {=} [A(\theta)x_0\!\!+\!\!B(\theta)u_0, \dots, A(\theta)x_{T-1}\!\!+\!\!B(\theta)u_{T-1}]^\top.
\end{gathered}
\label{eqn:kernel.vect}
\end{equation}

Following the main result in \cite[Theorem 1]{donati2025kernelTAC}, the residual function $\delta^\star$ minimizing \eqref{eq:kernel.problem} corresponds to the solution of 
\begin{equation}
\delta^\star  = \arg\min_{\delta\in\mathcal{H}}\sum_{t=0}^{T-1} \left[\tilde x_{t+1} - \delta(x_t)\right]^2 + \gamma\|\delta\|^2_{\mathcal{H}},
\label{eqn:thm1.dstar}
\end{equation}
with $\tilde x_{t+1} \doteq x_{t+1} - A(\theta)x_t-B(\theta)u_t$. Thus, it admits the representation
\begin{equation}
    \delta^\star(x) = \sum_{j=1}^T \omega_j^\star \kappa(x, x_j), \quad \omega_j^\star = \omega_j(\theta^\star),
    \label{eqn:delta.repr}
\end{equation}
where the coefficient vector is given by 
\begin{equation}
\omega(\theta) = (K + \gamma I_T)^{-1} (X - \Gamma(\theta)).
\label{eqn:omega.def}
\end{equation}
The optimal parameter vector $\theta^\star$ is thus obtained by solving the reduced optimization problem
\begin{equation}
    \begin{aligned}
    \theta^\star \!=\! \arg\min_{\theta \in \Theta} 
    &\sum_{t=0}^{T-1} \left[ x_{t+1} {-} A(\theta)x_t {-} B(\theta)u_t {-} K_t^\top \omega(\theta) \right]^2  \\&+ \gamma \omega(\theta)^{\top} {K}\omega(\theta),
    \end{aligned}
    \label{eq:reduced.kernel.problem}
\end{equation}
where $K_t^\top$ denotes the $t$-th row of the kernel matrix $K$.

Moreover, when \eqref{eqn:lure.sys} is affine in the parameters, so that, for the scalar case $n_x=1$, $n_u=1$, it can be written as 
$$\begin{aligned}
x_{t+1} &= A_0x_t + A^\top\theta x_t + B_0u_t
 + B^\top\theta u_t
 + F\phi(Cx_t)+v_t,\\
y_t &= Cx_t+w_t,
\end{aligned}$$
with  $A_0,B_0\in\mathbb{R}$, $A, B\in\mathbb{R}^{n_\theta}$,
the identification problem \eqref{eq:reduced.kernel.problem} is convex and its solution admits a closed-form expression, according to \cite[Theorem 2]{donati2025kernelTAC}.
To derive it, we define
$$
\begin{gathered}
X_0 = X - \left[ A_0 x_0 {+}B_0u_0, \dots, A_0 x_{T-1} {+}B_0u_{T-1} \right]^\top,\\ 
\Xi \!=\! \Bigl[A x_0 + B u_0, \dots, Ax_{T-1} + Bu_{T-1} \Bigr]^\top,
\end{gathered}
$$ $X_0\in\mathbb{R}^T$, $\Xi\in\mathbb{R}^{T,n_\theta}$, and
$$
\Psi \doteq (K + \gamma I_T)^{-1} \in\mathbb{R}^{T,T}.
$$
Then, the optimal parameter vector is given by:
\begin{equation}
    \theta^\star = \left( \Xi^\top \Psi \Xi \right)^{-1} \Xi^\top \Psi X_0,
    \label{eqn:reduced.kernel.linear}
\end{equation}
and the corresponding kernel coefficients in \eqref{eqn:delta.repr} are $$\omega^\star = (K + \gamma I_T)^{-1} (X_0 - \Xi \theta^\star).$$

\subsubsection{Extension to the general case}
We note that these results naturally extend to scenarios where $n_x > 1$ and $n_u > 1$, and full-state measurements are not available. In the latter case, the system output is given by $y_t = C x_t + w_t$, with $C \in \mathbb{R}$$^{n_y, n_x}$, and the nonlinearity is expressed as $\phi(C x_t)$. As detailed in \cite{donati2025kernelTAC}, a nonlinear smoother (such as the Unscented Rauch–Tung–Striebel smoother \cite{sarkka2008unscented}) can be applied to recover the latent state trajectory from the observed inputs and outputs. Once the state estimates are obtained, the kernel regression problem can be formulated and solved based on the reconstructed state-input pairs.

As for the multi-variable case, consider $n_x > 1$ and $n_u > 1$. Write the parametric matrices
$$
A(\theta)=\begin{bmatrix} a_1(\theta_1)^\top \\ \vdots \\ a_{n_x}(\theta_{n_x})^\top \end{bmatrix},\quad
B(\theta)=\begin{bmatrix} b_1(\theta_1)^\top \\ \vdots \\ b_{n_x}(\theta_{n_x})^\top \end{bmatrix},
$$
where $a_i(\theta_i)^\top\in\mathbb{R}$$^{1, n_x}$ and $b_i(\theta_i)^\top\in\mathbb{R}$$^{1, n_u}$ denote the $i$-th rows of $A(\theta)$ and $B(\theta)$, respectively. Here, we denote by $\theta_i$ the sub-vector of parameters that parametrize the $i$-th row. The nonlinear term $F\phi(x_t)$ decomposes componentwise $F\phi(x_t)$ accordingly, and can be written explicitly as
$
F\phi(x_t) =[
\Delta_1(x_t),
\dots,
\Delta_{n_x}(x_t)]^{\top},
$
where each component is given by
$
\Delta_i(x_t) = F_{i}^\top\phi(x_t)$, $i=1,\dots,n_x,$
with $F_{i}^\top\in\mathbb R$$^{1,n_\phi}$ denoting the $i$-th row of the matrix $F$.

It follows that the $i$-th state component satisfies the following scalar relation for each $i=1,\dots,n_x$, i.e.,
$$
x_{i,t+1} = a_i(\theta_i)\,x_t + b_i(\theta_i)\,u_t + \Delta_i(x_t) + v_{i,t}.
$$ 
For each $i$, the identification problem reduces to the scalar regression form treated in Section~\ref{sec:lure.kernel.method}: using the input pairs $(x_t,u_t)$  and the scalar outputs $x_{i,t+1}$, one can apply the same kernel-based estimation procedure to recover $\theta_i$ and an estimator $\delta_i$ of $F_i^\top\phi(x_t)$.

As these cases naturally extend from the full-state scalar setting, in the following we will present our approach in this simpler and more transparent scenario, while referring the reader to \cite{donati2025kernelTAC} for the extension to the general case.

\section{Lipschitz constant design via kernel regularization}\label{sec:lipdesign}
To ensure contractivity in the identified system, the learned model is required to be Lipschitz with a constant strictly smaller than one (see Definition~\ref{def:contract}). We now show how this condition can be enforced using a result from \cite{van2022kernel}, which characterizes the Lipschitz constant of kernel-based models in terms of the training data, the regularization parameter, and the chosen kernel.

As previously done, let $\kappa : \mathcal{Z}, \mathcal{Z} \to \mathbb{R}$ be a symmetric, positive semidefinite kernel, and let $\mathcal{H}$ denote its associated RKHS. Moreover, let us introduce the following definition.
\begin{definition}[Nonexpansive kernel]\label{def:nonexpansive.kernel}\itshape
    A kernel $\kappa : \mathcal{Z},\mathcal{Z} \to \mathbb R$ is nonexpansive if 
    $$
    |\kappa(z_1,z_1) - \kappa(z_1,z_2) - \kappa(z_2,z_1) + \kappa(z_2,z_2)|\!\leq\!\|z_1-z_2\|_{\mathcal{Z}}
    $$
    holds for all $z_1,z_2 \in\mathcal{Z}$, with $\|\cdot\|_\mathcal{Z}$ the norm in the input space $\mathcal{Z}$.
\end{definition}
Authors showed in \cite[Theorem~1]{van2022kernel} that, if a kernel $\kappa$ is \emph{nonexpansive} according to Definition~\ref{def:nonexpansive.kernel}, then every kernel function $g \in \mathcal{H}$ relating $\xi$ and $z$ as in \eqref{eqn:sample.sys} is Lipschitz continuous, i.e., 
$$
|g(z_1)-g(z_2)|\leq \ell_g\|z_1-z_2\|_2,
$$
and, in particular, the solution $\hat g \in \mathcal{H}$ to the regularized least-squares problem \eqref{eqn:reg.ls.prob.kernel}
has Lipschitz constant
\begin{equation}
\ell_g = \|\hat g\|_{\mathcal{H}} = \left\| \sqrt{K}(K + \gamma I_T)^{-1} Y \right\|_{2},
\label{eqn:kernel.lip.const}
\end{equation}
where $K$ denotes the positive semidefinite kernel matrix with entries $K_{ij} = \kappa(z_i, z_j)$, and $Y = [\xi_1, \dots, \xi_T]^\top$.

Consequently, the purely kernel-based function $\hat{g}$, solution of \eqref{eqn:reg.ls.prob.kernel}, has $\ell<1$ and is thus contractive according to Definition~\ref{def:contract}, if
$$
\left\| \sqrt{K}(K + \gamma I)^{-1} Y \right\|_2 < 1.
$$
This provides an explicit and testable condition to check the Lipschitz constants of kernel-based models, allowing the tuning of the regularization parameter $\gamma$ to design $\ell_g$, thereby satisfying the condition required for contractivity.

We note that the nonexpansiveness of the kernel itself is crucial to this result. Proposition~2 in \cite{van2022kernel} provides a practical criterion to ensure that a scalar-valued kernel $\kappa : \mathcal{Z} , \mathcal{Z} \to \mathbb{R}$ induces a nonexpansive operator-valued kernel. Specifically, \cite[Proposition~2]{van2022kernel} identifies concrete classes of kernels, such as Gaussian and Laplacian kernels, that are guaranteed to be nonexpansive according to Definition~\ref{def:nonexpansive.kernel}, meaning they induce operator-valued kernels whose associated RKHS contains only Lipschitz continuous operators.

We now show how to ensure that the identified model satisfies the Lipschitz condition required by contractivity, namely that the model map is contractive with a Lipschitz constant ${\mathcal{\ell}}<1$, exploiting the kernel-based approach first presented in \cite{donati2025kernelTAC} and resumed in the previous Section. We focus on Lur'e-type model structures approximating system \eqref{eqn:lure.sys}, of the form
\begin{equation}
	\hat x_{t+1} = A(\hat \theta)\hat x_t+ B(\hat\theta)u_t +  \delta(x_t).
    \label{eqn:lure.model}
\end{equation}
Here, $A(\hat\theta)$ and $B(\hat\theta)$ are the parametric matrices defining the linear dynamics. The estimated parameter vector, $\hat\theta$, is obtained by solving \eqref{eq:reduced.kernel.problem} or \eqref{eqn:reduced.kernel.linear}. On the other hand, $\delta$ is the kernel-based approximation of the nonlinear residual dynamics $\phi(Cx_t)$ in \eqref{eqn:lure.sys}, obtained via \eqref{eqn:delta.repr} as detailed in Section \ref{sec:lure.kernel.method}.

Due to the additive Lur’e-type structure, the Lipschitz constant of the model can be upper bounded in terms of the norm of its linear component and the Lipschitz constant of the nonlinearity, as formalized in the following lemma.
\begin{lemma}[Lur’e model Lipschitz bound]~ \label{prop:lure.lip}\itshape
	Consider the Lur’e-type model \eqref{eqn:lure.model}. Assume $\delta$ is Lipschitz with Lipschitz constant $\ell_\delta$ and $u_t\equiv0$, $\forall t$. Then, the following inequality holds:
	\begin{equation}
			\|\hat x_{t+1}-\hat{x}'_{t+1}\|_2 \le (\|A(\theta)\|_2 + \ell_\delta)\|\hat x_{t}-\hat{x}'_{t}\|_2,
            \label{eqn:lure.bound.lip}
	\end{equation}
	i.e., \eqref{eqn:lure.model} is Lipschitz with Lipschitz constant $\ell\leq \|A(\theta)\| + \ell_\delta$.
\end{lemma}
\begin{proof}
Let $\hat x_{t+1}$ and $\hat x'_{t+1}$ be two solutions of the model \eqref{eqn:lure.model}
corresponding to the input states $\hat x_t$ and $\hat x'_t$, respectively. Considering the difference $\hat x_{t+1}- \hat x'_{t+1}$, according to \eqref{eqn:lure.model}, we can write
\begin{equation*}
\hat x_{t+1}-\hat x'_{t+1} = A(\theta)(\hat x_t-\hat x'_t) + \left(\delta(\hat x_t)-\delta(\hat x'_t)\right).
\end{equation*}
Taking the norm and applying the triangle inequality yields
\begin{equation*}
    \|\hat x_{t+1}-\hat x'_{t+1}\|_2
\le \|A(\theta)(\hat x_t-\hat x'_t)\|_2 {+} \left\|\delta(\hat x_t)-\delta(\hat x'_t)\right\|_2.
\end{equation*}
Using the submultiplicativity of the induced $2$-norm and the Lipschitz property of $\delta$ (with constant $\ell_\delta$) we obtain
\begin{equation*}
\|\hat x_{t+1}-\hat x'_{t+1}\|_2
\le \|A(\theta)\|_2\|\hat x_t-\hat x'_t\|_2 + \ell_\delta\|\hat x_t-\hat x'_t\|_2,
\end{equation*}
which, collecting the terms, yields \eqref{eqn:lure.bound.lip}, concluding the proof. 
\end{proof}

Building on Lemma~\ref{prop:lure.lip}, which upper-bounds the model Lipschitz constant, we now give a simple sufficient condition that guarantees contractivity of the identified Lur’e-type model. The result follows immediately from the bound in Lemma~\ref{prop:lure.lip} and Definition~\ref{def:contract}.

\begin{theorem}[Lur'e model contractivity]\label{thm:contr}~ \itshape
Consider the Lur’e-type predictive model \eqref{eqn:lure.model} and a dataset $\mathcal{D}$ $= \{ (u_0,x_0),\dots , (u_{T-1},x_{T-1})\} \cup \{x_T\}$ collected from \eqref{eqn:lure.sys}. Define $X$ and $\Gamma(\theta)$ according to \eqref{eqn:kernel.vect}. Let $\delta$ be defined as in \eqref{eqn:delta.repr} via a nonexpansive kernel $\kappa: \mathcal{X, X} \to\mathbb R$, according to Definition~\ref{def:nonexpansive.kernel}. Given $K$ its positive semidefinite kernel matrix with entries $K_{ij} = \kappa(x_i, x_j)$, let $\theta^\star$ be defined as in \eqref{eq:reduced.kernel.problem} (or \eqref{eqn:reduced.kernel.linear}, when affine in $\theta$), with regularization weight $\gamma$. Then, the following two statements hold:
\begin{enumerate}
    \item[(i)] $\delta$ is Lipschitz in its state argument with constant \begin{equation}
    \ell_\delta = \left\| \sqrt{K}(K + \gamma I_T)^{-1} (X-\Gamma(\theta^\star)) \right\|_{2}.
    \label{eqn:lip.const.kern}
    \end{equation}
    \item [(ii)] If
\begin{equation}
\|\sqrt{K}(K {+} \gamma I_T)^{-1} (X {-} \Gamma(\theta^\star))\|_2 < {1{-}\|A(\theta^\star)\|_2},
\label{eqn:nonlin.cond}
\end{equation}
then the identified model \eqref{eqn:lure.model} with $\delta$ and $\theta^\star$ is strongly contractive with respect to the $\|\cdot\|_2$ norm, according to Definition~\ref{def:contract}. The contraction factor satisfies
\begin{equation}
\ell \le \|A(\theta^\star)\|_2 + \ell_\delta < 1.
\end{equation}
\end{enumerate}
\end{theorem}

\begin{proof}
Consider $\delta$ as in \eqref{eqn:delta.repr}, obtained from a nonexpansive kernel $\kappa$ satisfying Definition~\ref{def:nonexpansive.kernel}. By \cite[Theorem~1]{van2022kernel}, every function in the associated RKHS is Lipschitz continuous. In particular, $\delta$ is Lipschitz with constant given by \eqref{eqn:kernel.lip.const}. Since $\delta$ is the solution of \eqref{eqn:thm1.dstar}, we can express the regression residual as $Y \doteq (X - \Gamma(\theta))$, which allows us to rewrite \eqref{eqn:kernel.lip.const} in the equivalent form \eqref{eqn:lip.const.kern}, proving the first point of the claim.

To prove the second point, recall Lemma~\ref{prop:lure.lip}, which provides the Lipschitz bound~\eqref{eqn:lure.bound.lip} for the model in~\eqref{eqn:lure.model}. 
Assume~\eqref{eqn:nonlin.cond} holds and consider $\ell_\delta$ defined as in~\eqref{eqn:lip.const.kern}. Then $$
\begin{aligned}
\|A(\theta^\star)\|_2 + \ell_\delta
&< \|A(\theta^\star)\|_2 + {1-\|A(\theta^\star)\|_2}=1,
\end{aligned}
$$
and by Definition~\ref{def:contract} strong contractivity follows. This concludes the proof.
\end{proof}
Theorem~\ref{thm:contr} shows that contractivity of the identified Lur'e model can be guaranteed through a simple, explicit inequality that couples the linear gain $\|A(\theta^\star)\|_2$ with the Lipschitz constant of the kernel residual.  
This characterization is particularly useful from an identification viewpoint: it translates a global dynamical requirement into a verifiable algebraic condition directly computable from data.  
Moreover, the result reveals how regularization (via $\gamma$) and the linear parameters jointly influence contractivity, providing a clear mechanism to tune the model’s dynamical behavior.

\section{The proposed approach}\label{sec:prop_approach}
Building on this result, we propose two identification strategies that enforce condition~\eqref{eqn:nonlin.cond} through a joint tuning of the model parameters and the kernel regularization weight $\gamma$. In both approaches, the algorithm iterates over a logarithmically spaced grid of candidate regularization parameters $\mathcal{G}_\gamma$, while differing in how the contractivity condition is treated: checked {a posteriori} in Approach~\ref{sec:postcheck} and enforced as a constraint in Approach~\ref{sec:constrcheck}. 

\subsection{Unconstrained identification with post-contractivity check}\label{sec:postcheck}
In the first approach, 
for a fixed value of the regularization parameter~$\gamma\in\mathcal{G}_\gamma$, the parametric component~$\theta$ is identified via the kernel-based regression framework, i.e., solving \eqref{eq:reduced.kernel.problem} or via \eqref{eqn:reduced.kernel.linear} if in an affine-in-parameters setting, obtaining~$\theta^\star$. Then, the Lipschitz constant of the resulting model is evaluated according to \eqref{eqn:lip.const.kern}. Thus, the contractivity requirement \eqref{eqn:nonlin.cond} is checked. If it is met, the current value of $\gamma$ is stored in a feasible set. A new value of~$\gamma$ is then selected, and the procedure is repeated. In this way, the contractivity condition \eqref{eqn:nonlin.cond} is iteratively checked for all $\gamma \in \mathcal{G}_\gamma$. At the end of the loop, a final $\gamma^\star$ is chosen from the feasible set according to a pre-defined criterion (e.g., the one yielding the smallest validation RMSE). The solution corresponding to this selected value is then taken as the final identified model.

\subsection{Constrained identification with embedded contractivity check}\label{sec:constrcheck}
An alternative implementation can be formulated by incorporating the contractivity condition \eqref{eqn:nonlin.cond} directly within the optimization problem \eqref{eq:reduced.kernel.problem}, rather than checking it {a posteriori} over a grid of regularization parameters.
Specifically, considering \eqref{eqn:kernel.vect}, \eqref{eqn:omega.def} and \eqref{eq:reduced.kernel.problem}, the following optimization problem embeds condition \eqref{eqn:nonlin.cond} as a constraint within the optimization, i.e.,
\begin{equation}
\label{eq:constrained_J}
\begin{aligned}
    \min_{\theta \in \Theta} 
    &\sum_{t=0}^{T-1} \left[ x_{t+1} \!-\! A(\theta)x_t \!-\! B(\theta)u_t \!-\! K_t^\top \omega(\theta) \right]^2  \\&+ \gamma \omega(\theta)^{\top} {K}\omega(\theta),\\
\text{s.t. }
&\|A(\theta)\|_2
{+}\left\|\sqrt{K}(K{+}\gamma I_T)^{-1}(X {-} \Gamma(\theta))\right\|_2
{\le}1{-}\varepsilon.
    \end{aligned}
\end{equation}
Here, the scalar $\varepsilon>0$ defines a prescribed contractivity margin, ensuring that the estimated linear component $A(\theta)$ and the nonlinear kernel residual jointly satisfy a strict contraction condition. 
Conceptually, this approach replaces the post-check procedure of Section~\ref{sec:postcheck} with a single constrained solve that enforces the contractivity margin by construction. 
The resulting solution can be interpreted as the {projection} of the unconstrained estimate obtained from \eqref{eq:reduced.kernel.problem} or \eqref{eqn:reduced.kernel.linear} onto the set of contractive parameter values satisfying the contraction inequality (if any), ensuring that the identified predictor satisfies the contraction condition of Theorem~\ref{thm:contr}. 

If problem \eqref{eq:constrained_J} is feasible, the current value of~$\gamma$ is stored in the feasible set~$\mathcal{F}_\gamma$, and the procedure continues with the next element of~$\mathcal{G}_\gamma$. 
At the end of the loop, a final~$\gamma^\star$ is selected from~$\mathcal{F}_\gamma$ according to a predefined criterion, and the corresponding solution is taken as the final identified model.

Both procedures yield an identified model that explains the data and is guaranteed to be contractive, and are summarized in Algorithm~\ref{alg:alg1}.

\begin{algorithm}[!tb]
\caption{Identification of contractive Lur’e systems}
\label{alg:alg1}
\begin{algorithmic}[1]
\REQUIRE Dataset $\mathcal{D}=\{(u_0,x_0),\dots,(u_{T-1},x_{T-1})\}\cup\{x_T\}$ from \eqref{eqn:lure.sys}; nonexpansive kernel $\kappa$; margin $\varepsilon$; select approach $\mathsf{mode}\in\{\text{\ref{sec:postcheck},\ref{sec:constrcheck}}\}$.
\STATE Construct $X$, $\Gamma(\theta)$ as in~\eqref{eqn:kernel.vect} and the kernel matrix $K$ with $K_{ij}=\kappa(x_i,x_j)$.
\STATE Define a logarithmically spaced grid of candidates $\mathcal{G}_\gamma$ and set feasible set $\mathcal{F}_\gamma\gets\emptyset$.
\STATE \textbf{for each} $\gamma\in\mathcal{G}_\gamma$ \textbf{do}
\STATE \quad \textbf{if} $\mathsf{mode}=$ \ref{sec:postcheck} \textbf{then}
\STATE \qquad Identify $\theta^\star(\gamma)$ by solving~\eqref{eq:reduced.kernel.problem} (or with~\eqref{eqn:reduced.kernel.linear}).
\STATE \qquad Build $\delta(\gamma)$ via~\eqref{eqn:delta.repr} and compute $\ell_\delta(\gamma)$ using~\eqref{eqn:lip.const.kern}.
\STATE \qquad Compute and store a validation score (e.g., RMSE).
\STATE \qquad \textbf{if} the contractivity test~\eqref{eqn:nonlin.cond} holds \textbf{then} 
\STATE \qquad\quad $\mathcal{F}_\gamma {\gets} \mathcal{F}_\gamma {\cup} \{\gamma\}$.
\STATE \quad \textbf{else} {($\mathsf{mode}=$ \ref{sec:constrcheck})}
\STATE \qquad Solve the constrained problem~\eqref{eq:constrained_J} at this $\gamma$ (with\\\qquad prescribed margin $\varepsilon$) to obtain $\theta^\star(\gamma)$.
\STATE \qquad \textbf{if} problem~\eqref{eq:constrained_J} is feasible \textbf{then}
\STATE \qquad\quad Build $\delta(\gamma)$ via~\eqref{eqn:delta.repr}.
\STATE \qquad\quad Compute and store a validation score.
\STATE \qquad\quad $\mathcal{F}_\gamma {\gets} \mathcal{F}_\gamma {\cup} \{\gamma\}$.
\STATE \qquad \textbf{end if}
\STATE \quad \textbf{end if}
\STATE \textbf{end for}
\STATE \textbf{if} $\mathcal{F}_\gamma=\emptyset$ \textbf{then} report infeasible (no contractive solution\\\quad found on the grid).
\STATE \textbf{else} select $\gamma^\star\in\mathcal{F}_\gamma$ by the best validation score and set \\\quad$(\theta^\star,\delta)\gets(\theta^\star(\gamma^\star),\delta(\gamma^\star))$.
\STATE \textbf{Output:} Contractive identified model \eqref{eqn:lure.model} with parameters $\theta^\star$ and residual $\delta$.
\end{algorithmic}
\end{algorithm}

\section{Numerical example}\label{sec:examples}


We consider a contractive discrete-time Lur’e-type system of the form \eqref{eqn:lure.sys} obtained from the interconnection of a linear time-invariant (LTI) dynamics with a static nonlinearity. The contractive nature of this system was both proposed and validated in \cite{shima_contractivity_2025}. The linear part is defined by a state-space representation of dimension $n_x=3$, $n_u=1$, $n_\phi=1$ with matrices
\begin{equation}
\label{eqn:ex_sys}
A(\theta) =
\begin{bmatrix}
0 & \theta_1 & \theta_2 \\
\theta_3 & \theta_4 & 0 \\
0 & \theta_5 & \theta_6
\end{bmatrix}, 
\,
B = \begin{bmatrix} 0.1\\ 0.1\\ 0.2 \end{bmatrix},
\,
F = \begin{bmatrix} -0.2 \\ 0 \\ 0.2 \end{bmatrix},
\end{equation}
with $\theta_\mathrm{true} = [\theta_1,\dots,\theta_{n_\theta}]^\top {=} [-0.12, 0.3, 0.1, 0.8, 0.1, 0.6]^\top$ and $n_\theta=6$.
The feedback nonlinearity is introduced through a smooth static mapping of the form
$
\phi(x) = 0.5 \cos(0.5\,x_1)\sin(x_2),$ while the input signal $u_t$ is generated as a combination of sinusoidal excitations and additive Gaussian noise. The system is simulated over a horizon of $50$ time steps, starting from a random initial condition with entries sampled uniformly from $[0,1]$. To emulate realistic conditions, Gaussian perturbations with standard deviation $\sigma_d=0.01$ are added to the measured state. Hence, the available dataset containing input-output identification data is given by $\{(u_t, x_t), x_{t+1}\}$, ${t=0,\dots,T-1}$.

We first apply the kernel-based identification algorithm in Section~\ref{sec:postcheck}, using a Gaussian kernel of the form $k(x,x') = \exp\!\left(-\|x - x'\|_2^{2} / (2\sigma^{2})\right)$ with bandwidth $\sigma = 1$.  
The regularization parameter~$\gamma$ is tuned over a logarithmic grid, and for each candidate value the contractivity condition~\eqref{eqn:nonlin.cond} is verified. Specifically, we defined a logarithmically spaced grid $\mathcal{G}_\gamma$ containing values between  $\gamma_{min}=10^{-1}$ and $\gamma_{max}=10^3$. For each value of $\gamma\in\mathcal{G}_\gamma$, we evaluate the parametric identification error $E_\theta\doteq\|\theta^\star-\theta_\mathrm{true}\|_2$. The results are reported in Figure~\ref{fig:contrreg}, where the vertical axis represents the parametric error and the horizontal axis the regularization weight $\gamma$. The shaded region highlights the range of $\gamma$ for which the contractivity condition is satisfied, i.e., the set of feasible $\gamma$, $\mathcal{F}_\gamma$.
\begin{figure}
    \centering
    \includegraphics[width=0.9\linewidth]{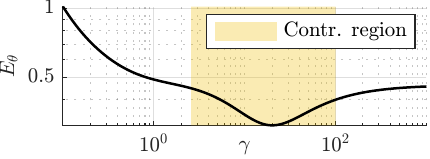}
    \caption{Parametric identification error and the contractive region with respect to $\gamma$.}
    \label{fig:contrreg}
\end{figure}
The figure clearly shows that enforcing contractivity yields lower parametric error, while non-contractive models (outside the shaded region) lead to poorer identification accuracy with respect to the parameters. 
%
Clearly, several aspects can guide the choice of $\gamma$ and, consequently, of $\ell_\delta$, within the contraction region. A systematic development of principled selection criteria will be the subject of future research.

Within this context, it is important to stress that the minimum prediction error (e.g., RMSE) does not necessarily correspond to the most accurate estimation of the physical parameters $\theta$ (see, e.g, Figure~\ref{fig:rmse}). 
\begin{figure}[!tb]
    \centering
    \includegraphics[width=0.9\linewidth]{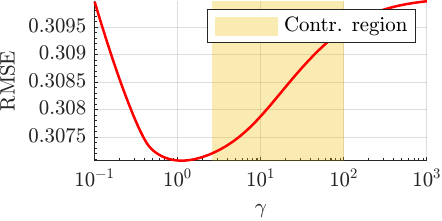}
    \caption{Root mean square errors with respect to $\gamma$.}
    \label{fig:rmse}
\end{figure}
This discrepancy arises because prediction errors are affected not only by the intrinsic system dynamics but also by external factors such as measurement noise, unmodeled disturbances, and stochastic variability in the data. As a result, a model that minimizes RMSE may do so by compensating for these external effects, at the cost of introducing a bias in the estimated parameters. In contrast, the ultimate goal of system identification is to recover the best possible parameter vector $\theta$ consistent with the underlying physics. Enforcing physical properties, such as contractivity, plays a key role in mitigating this bias, as it restricts the solution space to models that remain faithful to the structural properties of the true system. From this perspective, a slightly higher RMSE is not necessarily undesirable, provided that it reflects the unavoidable impact of noise and unmodeled effects, rather than inaccuracies in parameter estimation. This highlights how embedding physical insights into the identification procedure not only ensures desirable dynamical properties but also yields more reliable and interpretable parameter estimates, which are the primary objective of the identification task.
By jointly analyzing Figures~\ref{fig:contrreg} and~\ref{fig:rmse}, it becomes evident that the minimum value of the RMSE does not coincide with the minimum parametric error. However, a very small increase in the prediction error performance ($\approx 0.001$ higher) can correspond to a significantly lower error in the estimation of the parameter vector $\theta$ ($\approx 0.2$ lower), when contractivity requirements are satisfied.

To further assess the robustness of the proposed framework, we also evaluate the identification performance obtained using the approach described in Section~\ref{sec:constrcheck}, while keeping the underlying dynamical system~\eqref{eqn:ex_sys} unchanged. The feedback nonlinearity is modeled as the smooth static mapping 
$
    \psi(x) {=} \tfrac{1}{10}\log\!\left(e^{5x_2}{+}e^{-5x_2}\right){+}7
$ \cite{shima_contractivity_2025},
and the input sequence $u_t$ is generated following the same procedure adopted in the previous experiments.
The experiment follows the same protocol adopted in Algorithm~\ref {alg:alg1}: for a logarithmically spaced grid of~$\gamma$ values, the identification problem \eqref{eq:constrained_J} is solved. Multi-step prediction RMSE on a dedicated validation segment, constructed by splitting the simulated trajectory into a $70\%$ training portion and a $30\%$ validation portion, is used to determine the optimal regularization level.  
The kernel employed throughout the experiment is a Laplacian kernel of the form 
$k(x,x') = \exp(-\|x-x'\|_1 / \sigma)$ with bandwidth parameter $\sigma = 100$, evaluated on the state components.  

The results obtained for the most representative values of~$\gamma$ are reported in Table~\ref{tab:approachB_gamma_sweep_reduced}.
\begin{table}[!tb]
\centering
\footnotesize
\setlength{\tabcolsep}{5pt}
\caption{Representative identification results for Approach~\ref{sec:constrcheck}}
\label{tab:approachB_gamma_sweep_reduced}
\begin{tabular}{ccccccc}
\toprule
$\gamma$ & Feas. & $\|A\|_2$ & $\ell_\delta$ & $\|\theta^\star\!-\!\theta_\mathrm{true}\|_2$ & Val. RMSE \\
\midrule
$1.00\!\cdot\!10^{-4}$ & no  & --     & --      & --     & -- \\
$4.13\!\cdot\!10^{-3}$ & no  & --     & --      & --     & -- \\
\midrule
$1.43\!\cdot\!10^{-2}$ & yes & 0.7772 & 0.2218 & 0.0818 & 0.0133 \\
$1.61\!\cdot\!10^{-2}$ & yes & 0.7904 & 0.2086 & 0.0698 & 0.0127 \\
$1.82\!\cdot\!10^{-2}$ & yes & 0.8036 & 0.1954  & 0.0592 & \textbf{0.0125} \\
$2.32\!\cdot\!10^{-2}$ & yes & 0.8208 & 0.1739  & 0.0482 & 0.0140 \\
$5.88\!\cdot\!10^{-1}$ & yes & 0.8719 & 0.0809  & 0.1686 & 0.0913 \\
$1.00\!\cdot\!10^{3}$  & yes & 0.9884 & 0.0106 & 0.9694 & 0.4460 \\
\bottomrule
\end{tabular}
\end{table}
Feasibility is not attained for very small values of~$\gamma$, after which a stable feasible region emerges. The best validation performance is achieved at $\gamma = 1.82\!\cdot\!10^{-2}$, yielding a validation RMSE of $0.0125$, with similarly low errors in the neighbouring interval $\gamma\in[1.61\!\cdot\!10^{-2},\,2.32\!\cdot\!10^{-2}]$. Throughout this range, the identified models satisfy the contractivity requirement and exhibit reliable prediction accuracy.  Overall, the results highlight a clear trade-off between model smoothness (promoted by larger~$\gamma$) and the ability to capture unmodeled dynamics through the residual term.

The obtained results are then compared with a kernel estimator obtained via the closed-form expression~\eqref{eqn:reduced.kernel.linear} without imposing contractivity (Kernel-only).  
A summary of the performance metrics for the two models is reported in Table~\ref{tab:comparison_three_models}.
\begin{table}[tb]
\centering
\footnotesize
\setlength{\tabcolsep}{6pt}
\caption{Comparison of kernel models with and without contractivity constraint.}
\label{tab:comparison_three_models}
\begin{tabular}{lcc}
\toprule
{Metric} & 
Approach~\ref{sec:constrcheck} & 
{Kernel-only~\eqref{eqn:reduced.kernel.linear}}\\
\midrule
$\gamma^\star$ & $1.819\cdot10^{-2}$ & $4.125\cdot10^{-3}$ \\
Val. RMSE & $0.01250$ & $0.00918$ \\
$\|A\|_2$ & $0.8036$ & $0.8278$ \\
$\ell_\delta$ & $0.1954$ & $0.3671$ \\
$\|\theta^\star\!-\!\theta_\mathrm{true}\|_2$ & $0.0592$ & $0.1014$ \\
Contractivity & yes & no \\
\bottomrule
\end{tabular}
\end{table}
The closed-form kernel estimator without contractivity yields the best model at $\gamma=4.125\cdot10^{-3}$, slightly improving predictive accuracy over the contractive kernel fit. 
However, this unconstrained solution violates the contractivity condition ($\|A\|_2{+}\ell_\delta>1$) and exhibits a larger parametric error with respect to the kernel-based contractive model. 
This comparison further confirms that incorporating physical constraints during identification is not merely a robustness requirement but actively improves the fidelity and interpretability of the identified dynamics.


\section{Conclusions}\label{sec:conclusions}
In this paper, we have proposed a kernel-based framework for the identification of contractive Lur’e-type systems. By combining prior linear structure with a kernel representation of the nonlinear feedback, the approach enables accurate modeling while systematically enforcing contractivity. This ensures that the identified models not only fit the observed data but also preserve the fundamental properties of the underlying system.
Numerical experiments confirm that embedding contractivity into the identification procedure leads to improved parameter estimation and physically consistent models. These results highlight the importance of incorporating structural properties into identification procedures, particularly for applications where stability guarantees are essential.

Future work will focus on further investigation and testing of the proposed framework across different classes of nonlinear systems. Moreover, the Lipschitz-based design will be exploited to improve other identification criteria, such as the simulation error, and to leverage kernel methods for inferring additional system properties, for instance, through the use of Integral Quadratic Constraints (IQCs).

{
\bibliographystyle{IEEEtran}
\bibliography{root.bib}
}

\end{document}